\documentclass{article}

\usepackage{arxiv}
\usepackage{natbib}
\usepackage[utf8]{inputenc} 
\usepackage[T1]{fontenc}    
\usepackage[hidelinks]{hyperref}       
\usepackage{xurl}
\usepackage{url}            
\usepackage{booktabs}       
\usepackage{amsfonts}       
\usepackage{nicefrac}       
\usepackage{microtype}      
\usepackage{cleveref}       
\usepackage{graphicx}
\usepackage{doi}
\usepackage[labelfont=bf]{caption}

\title{OpenAlex: Features, Advantages and Limitations of an Open Database for Retrieving and Analysing Scholarly Outputs}

\usepackage{authblk}

\newbox{\orcid}\sbox{\orcid}{\includegraphics[scale=0.06]{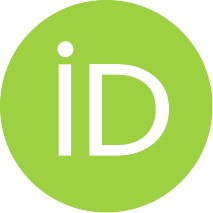}} 

\author{%
  \href{https://orcid.org/0000-0002-6462-3966}{\usebox{\orcid}\hspace{1mm}Ángel Borrego}%
}
\author{%
  \href{https://orcid.org/0000-0003-0935-6436}{\usebox{\orcid}\hspace{1mm}Cristóbal Urbano}%
}

\affil{Universitat de Barcelona\\
Facultat d’Informació i Mitjans Audiovisuals\\
Centre de Recerca en Informació, Comunicació i Cultura (CRICC)\\
Melcior de Palau, 140; 08014 Barcelona, Spain \\
\texttt{borrego@ub.edu; urbano@ub.edu} \\
}

\renewcommand{\shorttitle}{OpenAlex: features, advantages and limitations}

\hypersetup{
pdftitle={OpenAlex: Features, Advantages and Limitations of an Open Database for Retrieving and Analysing Scholarly Outputs},
pdfauthor={Borrego, Á., Urbano, C.},
}

\begin{document}
\maketitle

\begin{abstract}
OpenAlex is an open bibliographic database that has been proposed as an alternative to commercial platforms in a context defined by the aim of transforming science evaluation systems into more transparent sources based on open data. This paper analyses its features, information sources, entities, advantages and limitations. The results reveal numerous records lacking abstracts, affiliations and references; deficiencies in identifying document types and languages; and issues with authority control and versioning. Although OpenAlex has been adopted in important initiatives and has yielded results comparable to those obtained with commercial databases, gaps in its metadata and a lack of consistency point to a need for intensive data cleaning, suggesting it should be used with caution. The study concludes by identifying three lines of action to improve data quality: increasing publishers’ commitment to completing metadata in primary sources; creating coordination structures to channel the contributions of institutional users; and endowing the project with sufficient human resources and reliable procedures to address internal quality control tasks and user support requests.
\end{abstract}


\section{Introduction}

Citation indexes originated in 1963 with Eugene Garfield's creation of the Science Citation Index, initially covering 613 journals and 1.4 million citations \citep{Urbano2016Garfield}. This database not only provided an innovative information retrieval system based on citations that enabled the tracking of influences and connections between studies, but also revolutionised scientific evaluation by facilitating the measurement of a publication’s visibility. The coverage was expanded with the creation of the Social Sciences Citation Index and the Arts and Humanities Citation Index, which were eventually integrated into the Web of Science platform.

For years, Web of Science held an almost exclusive position in global scientific evaluation as the main source for citation analysis and impact measurement of academic publications. This dominance began to decline in 2004 with the launch of Scopus, a database developed by Elsevier with expanded journal coverage, especially in social sciences and humanities, as well as non-English publications \citep{Mongeon2016Coverage}.

Also in 2004, Google Scholar was launched as a free tool with a different approach, based on automated full-text indexing of a broad range of academic documents available on the web, including articles, books, reports and theses. Google Scholar offers much wider coverage than traditional bibliographic databases, although with less control over the quality of its sources \citep{OrdunaMalea2016GoogleScholar}.

In 2009, Microsoft Academic was launched to compete with Google Scholar. Although its early development was limited and it was temporarily suspended, the project was relaunched in 2016 with enhanced coverage and advanced semantic analysis capabilities. Despite these improvements, Microsoft announced\footnote{\url{https://www.microsoft.com/en-us/research/articles/microsoft-academic-to-expand-horizons-with-community-driven-approach/}} the definitive retirement of Microsoft Academic in May 2021 \citep{Chawla2021MAGDiscontinued}.

Shortly after Microsoft Academic was shut down, in early 2022, OurResearch announced the launch of OpenAlex.\footnote{\url{https://blog.ourresearch.org/were-building-a-replacement-for-microsoft-academic-graph/}} OurResearch is a nonprofit organization known for creating Unpaywall, a browser extension that indexes free and legal open access versions of scientific articles. OpenAlex—named after the Library of Alexandria—is an open bibliographic database that aims to provide a sustainable, accessible and open-source alternative for analysing global scientific output, drawing part of its model and data from Microsoft Academic Graph \citep{Chawla2022OpenIndex}. In September 2025, OurResearch announced a brand name change to OpenAlex,\footnote{\url{https://blog.openalex.org/were-now-openalex/}} encompassing both Unpaywall and Unsub, a tool for evaluating journal subscription packages and cancellation options.

OpenAlex is an open-source database, which means that its data is freely accessible to anyone who wishes to consult, analyse or reuse it. Unlike Web of Science or Scopus, OpenAlex allows its information to be consulted and downloaded without a subscription. Moreover, its technical infrastructure and data models are publicly documented, fostering transparency and the development of new tools \citep{Priem2022OpenAlex}.

OpenAlex is funded through grants, donations and premium clients. In February 2024, the French Ministry of Higher Education and Research announced financial support for OpenAlex,\footnote{\url{https://www.ouvrirlascience.fr/french-ministry-of-higher-education-and-research-partners-with-openalex-to-develop-a-fully-open-bibliographic-tool/}} recognising it a crucial infrastructure for open science and committing to improving its data, particularly in relation to French research. A month later, in March 2024, OpenAlex announced it was receiving a \$7.5 million grant from the Arcadia Foundation.\footnote{\url{https://blog.openalex.org/ourresearch-receives-7-5m-grant-from-arcadia-to-establish-openalex-a-milestone-development-for-open-science/}} At the same time, OpenAlex earns revenue from premium subscribers who receive more frequent database updates, unlimited API queries and priority support.\footnote{\url{https://help.openalex.org/hc/en-us/articles/24397762024087-Pricing}} OpenAlex thus operates as a nonprofit project sustained by donations and premium subscriptions. It is not yet clear whether this model will be sustainable in the long-term\footnote{The fact that OpenAlex CEO Jason Priem organised a webinar in May 2025 to present his view of the sustainability of the project is a clear indication that users are asking crucial questions about this issue. The recording is available on OurResearch’s YouTube channel: \url{https://youtu.be/CZ5Q9To1zCc?si=k7Pht-pz8Q0D8ky}} or will eventually cease to operate or even revert to the traditional commercial data model \citep{Cao2025AcademiaOpenData}.

OpenAlex is used for the CWTS Leiden Ranking Open Edition\footnote{\url{https://open.leidenranking.com}} and French institutions such as Sorbonne University\footnote{\url{https://www.sorbonne-universite.fr/en/news/sorbonne-university-unsubscribes-web-science}} and the Centre National de la Recherche Scientifique (CNRS)\footnote{\url{https://www.cnrs.fr/en/update/cnrs-has-unsubscribed-scopus-publications-database}} have announced its adoption as an alternative to Scopus and Web of Science. Interest in OpenAlex has grown notably in the wake of the Barcelona Declaration on Open Research Information\footnote{\url{https://barcelona-declaration.org}} issued in April 2024, which proposes the transformation of the scientific evaluation system by promoting the use of more transparent indicators aligned with open science principles. In this context, OpenAlex has been identified as a strategic alternative thanks to the free, auditable bibliographic data it provides.

\section{Sources of OpenAlex}

OpenAlex draws from a combination of open sources, most notably Microsoft Academic Graph and Crossref,\footnote{\url{https://help.openalex.org/hc/en-us/articles/24397285563671-About-the-data}} as well as recent significant additions such as DataCite.\footnote{\url{https://datacite.org/blog/datacite-metadata-is-now-integrated-in-openalex/}}

Crossref is a nonprofit organisation founded in 2000 to facilitate consistent identification and linking of online scientific content. Its primary role is to act as a DOI registration agency for academic publications. DOIs uniquely identify digital objects (articles, datasets, monographs, reports, etc.) and remain unchanged throughout the life of the document's life, linked to its metadata, including its URL. This ensures stable access to documents, provided publishers update metadata when URLs change.

Crossref is a key source for OpenAlex because it provides metadata on academic content registered by publishers: document title, authors, affiliations, journal, publication date, references, etc. As an open source, Crossref metadata can be freely reused. However, the accuracy and reliability of the metadata in Crossref depends on the information supplied by the publishers participating in the system, as the organisation does not perform cleaning or curation tasks on these data.\footnote{\url{https://www.crossref.org/documentation/principles-practices/}}

Some publishers may provide incomplete metadata, resulting in partial records that do not reflect all the relevant information on a publication. In some cases, the lack of verification controls can lead to the inclusion of erroneous or even false metadata, which affects the reliability of systems that depend on these data, such as OpenAlex. A study by \citet{Besancon2024Sneaked} revealed the inclusion in Crossref of "sneaked" references registered as metadata for scientific articles in which they did not actually appear, exploiting the trust relationship between publishers and Crossref. Their analysis of three journals from the same publisher found that at least 9\% of the references analysed were "clandestine". Although these citations did not appear in the published articles, they were deposited in Crossref and propagated to the records of platforms that use it as a data source.

\section{OpenAlex entities}

In OpenAlex, entities are the basic units that structure and organise bibliographic information within the database. OpenAlex uses nine types of entities: \textit{works}, \textit{authors}, \textit{sources}, \textit{institutions}, \textit{topics}, \textit{keywords}, \textit{publishers}, \textit{funders} and \textit{geo}.

The main characteristics of the first six entities are described below, using examples obtained through queries submitted to the web version of the database during the last two weeks of November 2025. The OpenAlex search interface, which \citet{Codina2024OpenAlexReview} describes as more limited and with fewer functionalities than Scopus or Web of Science, is not analysed here.

\subsection{Works}

These are individual academic publications, such as articles, books, datasets or reports. Each work includes information about the title, authors, source and publication date, as well as citation relationships with other works (cited references and citations received).

\begin{figure}[ht]
    \centering
    \includegraphics[width=\textwidth]{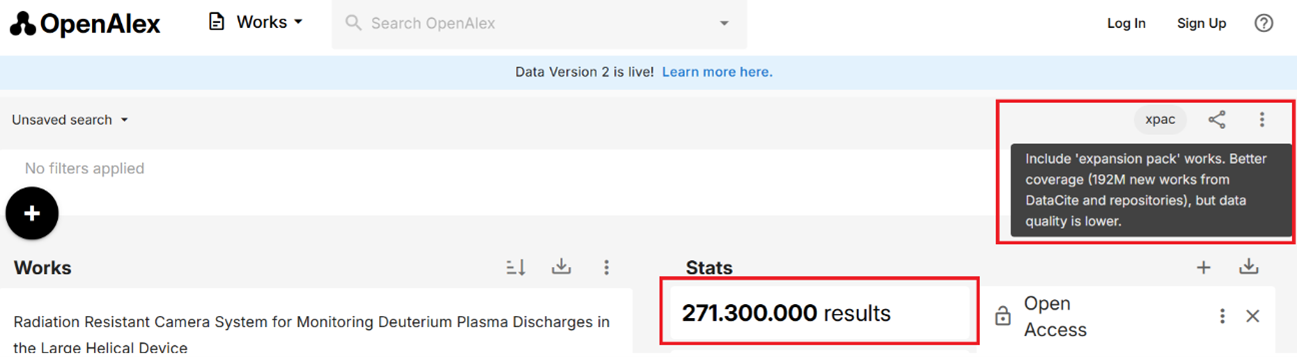}
    \caption{Works available in OpenAlex}
    \label{fig:works}
\end{figure}

As of November 2025, OpenAlex has 271.3 million indexed works, to which an "expansion pac" (\textit{xpac}) can be added containing another 192 million records with lower-quality metadata, sourced from DataCite and institutional repositories (\autoref{fig:works}).

\subsubsection{Abstracts}

In the standard version (without including the \textit{xpac}), 40\% of the records (107.3 million) lack an abstract.\footnote{\url{https://openalex.org/works?page=1&filter=has_abstract:false}} The presence of an abstract depends on whether the original source provides it and whether it can be legally disseminated. Not all publishers include abstracts in the metadata they submit to Crossref or other sources used by OpenAlex; in many cases, they only publish them on their own websites. Moreover, some abstracts are protected by copyright, which means OpenAlex cannot share them. A study by \citet{VanEck2025Crossref} on the deposit of abstracts in Crossref in 2023 and 2024 revealed that publishers such as Elsevier, Taylor \& Francis, IEEE and the American Chemical Society had not deposited any during the two-year period studied.

\subsubsection{Document types}

Identifying the document type of works in OpenAlex is often problematic. A study of a sample of 6.6 million records \citep{Mongeon2025Discrepancies} found more than 300,000 cases of publications classified as "article" or "review" in OpenAlex but assigned a different document type in Web of Science. Almost all manually verified cases pointed to an incorrect classification in OpenAlex. The discrepancies identified in this study did not include cases where Web of Science classified a text as an "article" while OpenAlex classified it as a "review", or vice versa. Including these discrepancies reveals a significantly higher level of divergence between the two sources.

A search in OpenAlex for documents with the phrase "new editor in chief" in the title retrieves 1,751 results.\footnote{\url{https://openalex.org/works?page=1&filter=display_name.search:new+editor+in+chief}} Although most of these texts could be expected to be editorials, only 20\% (350 records) are classified as such, while the majority (74\%, 1,301 records) are categorised as articles.

It is also debatable whether the inclusion of library guides (\textit{libguides}), which account for more than 1.7 million records,\footnote{\url{https://openalex.org/works?filter=type:types/libguides}} support an effective analysis of scientific activity, or to group a collection of more than 3.8 million records in the category of "paratext".\footnote{\url{https://openalex.org/works?filter=type:types/paratext}} On the other hand, patents are not included.

\subsubsection{Languages}

Determining the language of documents is also problematic. A study by \citet{Mongeon2025Discrepancies} analysed the accuracy of language identification by comparing English with other languages. Although some discrepancies were observed, they were less common than in the case of document types, and no clear difference was identified between Web of Science and OpenAlex.

A manual analysis of the completeness and accuracy of language metadata in a sample of 6,836 articles in OpenAlex \citep{Cespedes2025LinguisticCoverage} suggests that it offers more balanced language coverage than Web of Science, although errors were detected in identifying document languages, leading to an overestimation of the number of English documents in the database.

Both the abovementioned studies conclude that errors in the identification of the languages of works have two basic causes: mistakes made by OpenAlex’s language detection algorithm and the fact that some journals publish articles and abstracts simultaneously in multiple languages.

\subsubsection{References}

References constitute yet another metadata element that poses difficulties. In the standard version of OpenAlex (excluding \textit{xpac}), a search for documents without references (reference count = 0)\footnote{\url{https://openalex.org/works?page=1&filter=referenced_works_count:0}} retrieves 64\% of the records in the database (173.5 million). 

\citet{Culbert2025ReferenceCoverage} analysed 16.8 million recent publications indexed by three databases, concluding that OpenAlex contained average reference counts per source and internal coverage levels comparable to those of Web of Science and Scopus. However, while OpenAlex included 586 million references for these records, Web of Science had 725 million and Scopus 727 million. Similarly, \citet{Thelwall2025Suitability} concluded that OpenAlex is suitable for citation analysis in most fields. Another study by \citet{Scheidsteger2025Similarity} compared field-normalised citation indices obtained from OpenAlex with those calculated by three commercial databases: Web of Science, Scopus and Dimensions. The study focused on about 335,000 articles published between 2013 and 2017 by 48 German universities in four major subject areas. Although general agreement was observed at article level, significant differences were detected when the data were broken down by university and subject area.

When creating a work record, OpenAlex does not include references to documents that are not indexed in the database. The references cited in a document may be other documents indexed in the database or documents not indexed. In Web of Science or Scopus, references in the second category are added to the database as secondary entries: the documents are not retrieved in a query, but they are identified as citations. In OpenAlex, these references are lost.

The most cited work among those retrieved in a query for documents without references is "R: A Language and Environment for Statistical Computing",\footnote{\url{https://openalex.org/works/w2582743722}} a well-structured piece of grey literature that includes a "References" section listing nine documents that remain invisible in OpenAlex. This example illustrates OpenAlex's openness to a much broader range of document types than commercial databases, while also highlighting its inability to consistently process references for all documents.

\begin{figure}[ht]
    \centering
    \includegraphics[width=\textwidth]{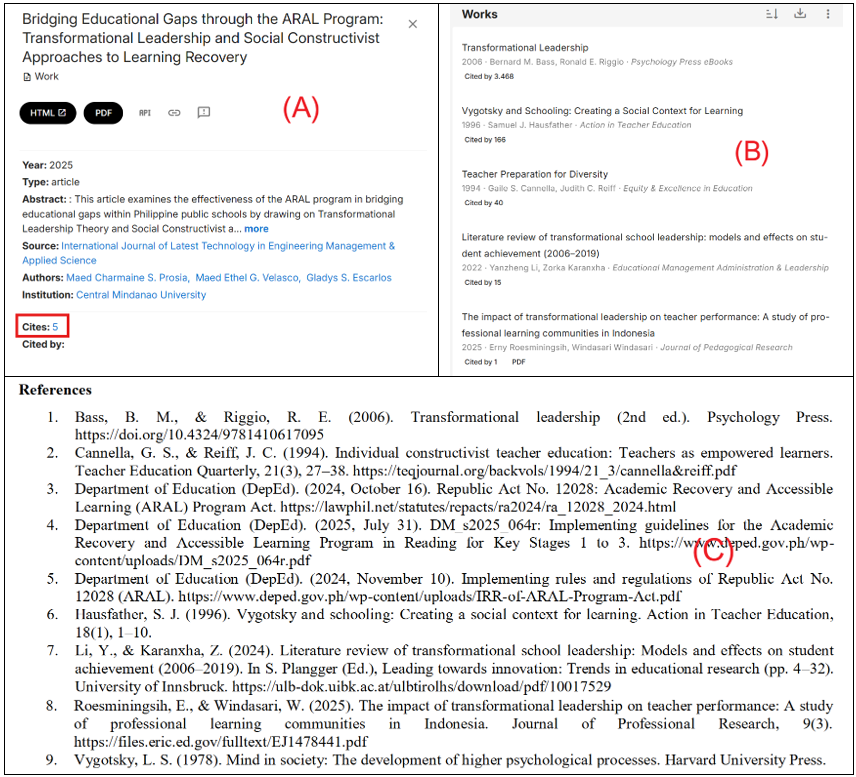}
    \caption{(A) Record in OpenAlex; (B) References in the OpenAlex record; (C) References in the original document}
    \label{fig:references}
\end{figure}

\autoref{fig:references} shows another example of this loss of references. The upper left section (A) contains a record for a document that OpenAlex identifies as having five references.\footnote{\url{https://openalex.org/works/w4416039086}} In the upper right section (B), these five references are listed.\footnote{\url{https://openalex.org/works?page=1&filter=cited_by:w4416039086}} Finally, the lower section (C) shows the nine references included in the original article.\footnote{\url{https://doi.org/10.51583/ijltemas.2025.1410000050}} Only references 1, 2, 6, 7 and 8 from the source document have been incorporated into the OpenAlex record. Three references from the U.S. Department of Education and one reference to a monograph (reference 9 in the original document), despite being indexed in OpenAlex, have been lost.\footnote{\url{https://openalex.org/works/w2135943618}}

\subsection{Authors}

\begin{figure}[ht]
    \centering
    \includegraphics[width=\textwidth]{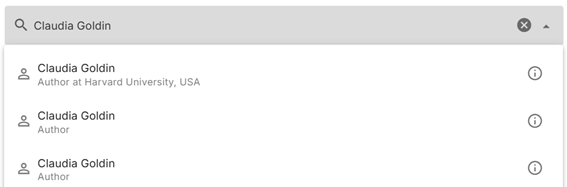}
    \caption{Claudia Goldin's profiles in OpenAlex}
    \label{fig:goldin}
\end{figure}

In OpenAlex, it is common to find authors with duplicate profiles. For example, on its homepage, OpenAlex suggests "Claudia Goldin", an American economist who received the Nobel Prize in 2023, as an example for an author search. However, a search for this author reveals that her publications are scattered across three separate profiles (\autoref{fig:goldin}).

OpenAlex provides a form to request corrections to author profiles,\footnote{\url{https://help.openalex.org/hc/en-us/articles/27283405287319-How-can-I-fix-errors-in-an-OpenAlex-author-profile}}  such as modifying the name format, merging multiple profiles for the same author, removing publications belonging to other authors, or adding missing publications. However, OpenAlex's responsiveness to these requests is currently very slow, so many remain in "open" status for months.\footnote{For example, a "request" submitted on 27 August 2025 (ID 5410) was still unresolved as of 30 November 2025. It is unclear whether premium users enjoy better response times.}

\subsection{Institutions}

A frequent issue in OpenAlex is the absence of institutional affiliations. According to a study by \citet{Zhang2024MissingInstitutions}, more than 60\% of records in OpenAlex lack institutional information entirely or partially. This problem is especially common among older publications and in the fields of social sciences and humanities.

Another problem with the information in this category is that some affiliations assigned to certain documents are incorrect. This issue has received less attention in the literature because it requires manual verification. \citet{Bordignon2024OpenAlex} examined how much of the bibliographic output of École des Ponts\footnote{A French engineering school belonging to the Institut Polytechnique de Paris. Website: \url{https://ecoledesponts.fr}} indexed in Web of Science, Scopus or HAL was also indexed in OpenAlex, and whether the publications retrieved when searching for this institution in OpenAlex were correctly assigned. The results highlighted the quality of OpenAlex's recall, as it retrieved 93\% of École des Ponts' output, but also its lack of precision: 24\% of the documents retrieved in searches for this institution belonged to other institutions.

\begin{figure}[ht]
    \centering
    \includegraphics[width=\textwidth]{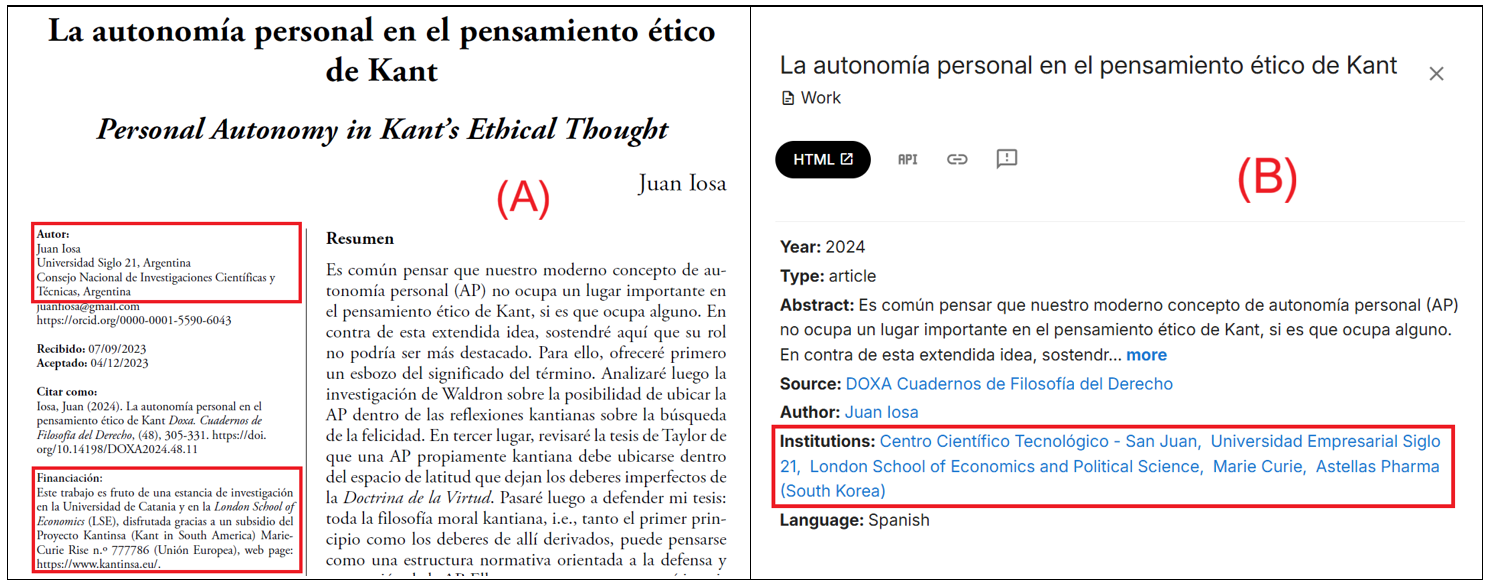}
    \caption{(A) Original document; (B) OpenAlex record showing affiliations from the funding section and others unrelated to the document}
    \label{fig:iosa}
\end{figure}

\autoref{fig:iosa} shows an example of an OpenAlex record\footnote{\url{https://openalex.org/works/w4402770154}} in which the algorithm combined the author's affiliation with institutions mentioned in the funding section\footnote{\url{https://doi.org/10.14198/DOXA2024.48.11}} and other organisations that appear unrelated to the document, such as a South Korean pharmaceutical company. These incorrect affiliations extend to author profiles, as it is common to find author profiles in OpenAlex with dozens of institutional affiliations, many of which the author has never been associated with.\footnote{This problem affects, for example, the authors of this preprint: for Ángel Borrego there are eight incorrect "past institutions" (\url{https://openalex.org/authors/a5079282617}); for Cristóbal Urbano, there are 11 (\url{https://openalex.org/authors/a5065832938}).}

In France, the Ministry of Higher Education and Research has launched the Works-magnet\footnote{\url{https://works-magnet.esr.gouv.fr}} initiative to "accelerate" \citep{Jeangirard2024WorksMagnet} the improvement of bibliographic metadata for French publications through review and validation by librarians and research managers. This initiative has been driven by the French community of users, as the governmental (the Ministry) and institutional (the CNRS and the Sorbonne) commitment to this tool has led library staff and research managers to analyse its reliability in relation to their institutions \citep{Bordignon2024OpenAlex,Bach2025OpenAlexExperience}. Community-driven initiatives like this one could pave the way for OpenAlex's growth as a collaborative ecosystem, which is why OpenAlex refers users—even from other countries who need to identify problems and propose solutions—to Works-magnet. However, despite Works-magnet’s role as an "accelerator", its current response capacity is limited by the sheer volume of errors to be corrected and the relatively limited resources available to the OpenAlex team, which ultimately must implement the required changes.\footnote{The changes requested via Works-magnet are recorded in the "Issues" section (\url{https://github.com/dataesr/openalex-affiliations/issues}) of the GitHub profile "openalex-affiliations", which "exhibits cases where OpenAlex affiliation could be improved." As of 30 November 2025, the "Issues" section lists 106,087 open cases, of which only 51,420 appear as closed. The backlog of pending cases is growing rapidly, as by the end of August 2025, 73,995 issues had been received and 49,903 had been closed.}

\subsection{Sources}

The sources where works indexed in OpenAlex are published include journals, conferences, repositories and databases. This diversity of sources poses challenges in controlling document versions, as primary sources that publish documents (such as journals) are combined with secondary sources that only publish references to documents published in primary sources (such as indexing and abstracting databases).\footnote{As an example of their heterogeneous composition, it is worth noting that the top ten sources by number of records are: PubMed, Medical Entomology and Zoology, arXiv, HAL, Dialnet, SSRN, Zenodo, Global Biodiversity Information Facility, DOAJ and Elsevier eBooks: \url{https://api.openalex.org/works?group_by=primary_location.source.id&per_page=10}} Other sources, such as repositories, mix primary documents with secondary documents, such as preprints or postprints of primary articles. In short, the diversity of indexed sources requires excellent version control for documents, which OpenAlex does not always achieve.

\begin{figure}[ht]
    \centering
    \includegraphics[width=\textwidth]{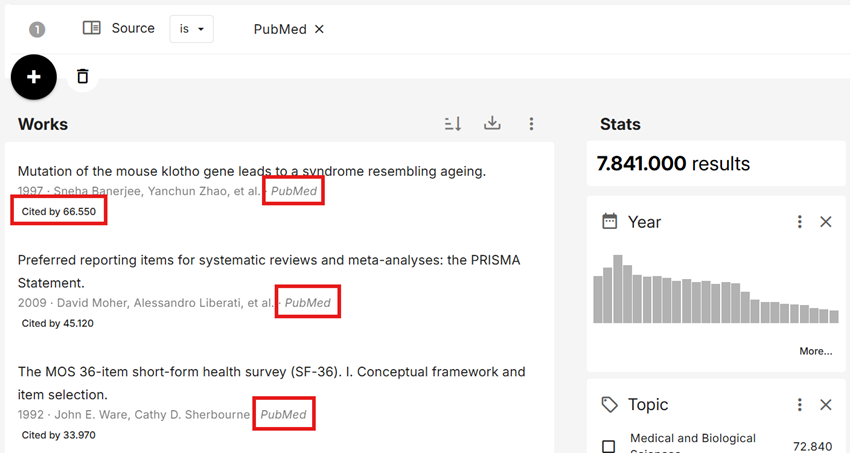}
    \caption{Results of a search for documents with PubMed as the source}
    \label{fig:pubmed}
\end{figure}

\autoref{fig:pubmed} shows the first records retrieved in a query for documents whose source is PubMed, sorted by citation count.\footnote{\url{https://openalex.org/works?page=1&filter=primary_location.source.id:s4306525036}} In fact, PubMed does not publish primary documents but only indexes articles published in other sources, mainly journals.

\begin{figure}[ht]
    \centering
    \includegraphics[width=\textwidth]{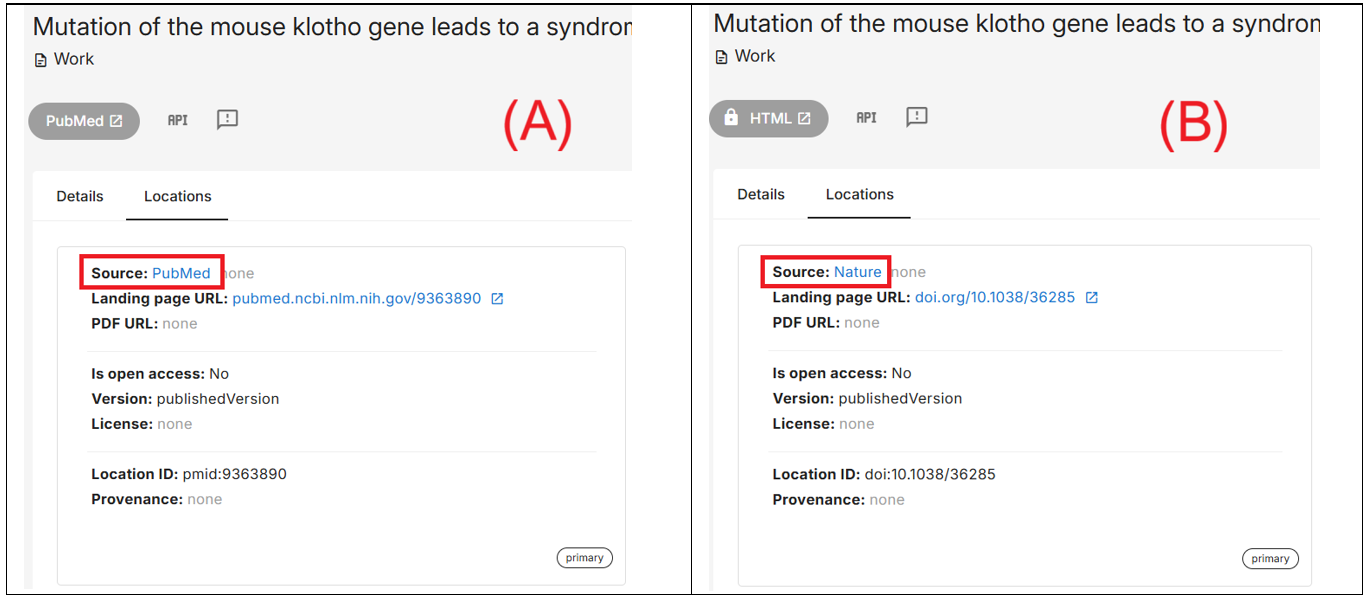}
    \caption{Versions of the same document identified as different records}
    \label{fig:versions}
\end{figure}

According to the first record shown in the query results, the version of the document in PubMed is the primary one (\autoref{fig:versions}A).\footnote{\url{https://openalex.org/works/w1964184380}} In reality, this article was originally published in \textit{Nature} and has its own record in OpenAlex, which should be linked to the PubMed record, with the version in \textit{Nature} defined as the primary version (\autoref{fig:versions}B).\footnote{\url{https://openalex.org/works/w2114360920}} 

Another notable aspect of the PubMed record is that OpenAlex identifies it as having 66,550 citations, whereas the \textit{Nature} record only shows 3,729, a figure similar to the 3,473 citations indicated on the journal’s website.\footnote{\url{https://doi.org/10.1038/36285}}

\subsection{Topics and keywords}

OpenAlex tags works with various topics using an automated system that considers the title, abstract, source and references. The topic with the highest score is the document's main topic. This main topic in turn belongs to a subfield, field and domain, which can be used to classify the work from the most specific to the most general level of granularity.

\begin{figure}[ht]
    \centering
    \includegraphics[width=0.7\textwidth]{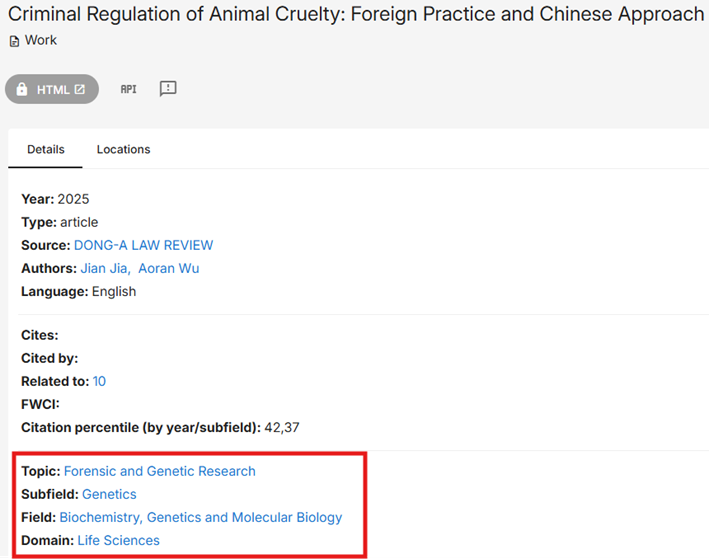}
    \caption{Thematic classification of an article on the regulation of animal cruelty}
    \label{fig:law}
\end{figure}

This classification system and the assignment of a single topic to each document limits thematic retrieval. For example, in a field such as law, which encompasses the study of regulatory aspects of a wide range of activities, it is common for the activity itself to be identified as the topic while the legal nature of the document is overlooked. This is the case of the article shown in \autoref{fig:law}, where the thematic classification does not include any reference to law despite being an article on the regulation of animal cruelty in China.\footnote{\url{https://openalex.org/works/w4414088320}}

OpenAlex assigns keywords to works based on the topics. These keywords do not appear in the record, even though they are available as a searchable field.

\section{Conclusions}

OpenAlex is a free and open database presented as an alternative to commercial platforms such as Web of Science and Scopus. Interest in its use has grown in a context where the scientific evaluation system is being transformed by the promotion of open sources that contribute to greater transparency in evaluation processes.

Although OpenAlex incorporates functionalities similar to those of commercial databases, it suffers from significant limitations in metadata quality. Among the most notable issues are the large volume of records lacking abstracts, affiliations and references, as well as deficiencies in identifying document types and languages. Added to these issues are problems with authority control for authors and institutions and with document version management, which results in the appearance of duplicates.

These results suggest that caution is needed when using OpenAlex. The database has been used in products such as the open edition of the CWTS Leiden Ranking, with results relatively consistent with those obtained in the edition based on Web of Science. However, the preparation of these rankings involves specific cleaning processes, such as the exclusion of works not published in English or in national journals \citep{VanEck2024CoreSources}, which restrict the analysis to a set of sources that are better represented in Web of Science and, at the same time, that usually provide more complete metadata to Crossref and other sources feeding OpenAlex.

Citation information for a much broader universe of documents than that offered by commercial databases is one of the reasons why OpenAlex plays a prominent role in the discourse on open sources for research evaluation, as consolidated in the Barcelona Declaration. OpenAlex's greater coverage of source documents, free access and availability of open data that can be reused without restrictions will lead to a proliferation of bibliometric studies and evaluation reports based on this database in the coming years. However, there is a need for greater clarity on the reliability of its citation counts, as in the absence of a systematic study, there is reason to believe that the examples identified here may be indicative of a pattern. \autoref{fig:versions}, for example, shows a significant discrepancy in the number of citations for the same document depending on which source used by OpenAlex is considered.

The most effective strategy to improve data quality would be to strengthen the completeness of primary sources, especially Crossref. However, this will require active commitment from publishers. In this regard, a study by \citet{Kramer2025CrossrefMetadata} notes that the ability of publishers to register metadata in Crossref largely depends on their capacity to capture and maintain it in their own production workflows. Manuscript management systems seem to play an important role in determining what information is effectively made available to Crossref. However, significant differences are also observable between publishers using the same system, suggesting that technology is not the only determining factor and that commercial considerations also play a role.

A complementary way to improve metadata would be for it to be completed by institutional users. However, it remains to be seen whether this option can be implemented in a decentralised manner given that there is no clearly defined community of users or established coordination mechanisms. While numerous institutions have signed declarations supporting more transparent evaluation processes based on open sources, it is important to remember that these initiatives also entail costs and require sustained resources.

A project with the objective stated by OpenAlex needs resources and processes in line with the challenge assumed. At present, OpenAlex's human team appears to be very small, while its processes based on automated and algorithmic document processing leave considerable room for improvement, as evidenced in the discussion above of the assignment of author affiliations. The data on response times and the backlog of unresolved issues reported in recent months, along with the expansion in the number of new sources used to increase the number of records, point to significant risks in relation to quality control of the information recorded.

\bibliography{references}

\end{document}